\documentclass[
aps,prl,
amssymb,
twocolumn,
floatfix,
superscriptaddress,
 amsmath,amssymb,
]{revtex4-2}

\usepackage{graphicx,subfigure}
\usepackage{dcolumn}
\usepackage{bm}

\usepackage{ulem}
\usepackage{comment}
\usepackage{color}
\usepackage[dvipsnames]{xcolor}

\newcommand{\eg}[1]
{\textcolor{red}{[eg: #1]}}

\newcommand{\ag}[1]
{\textcolor{brown}{[ag: #1]}}

\begin{document}


\title{
Unveiling Crystalline Order from Glassy Behavior
of Charged Rods
at Very Low 
Salt Concentrations}

\author{Hanna Anop}
\affiliation{Univ. Bordeaux, CNRS, Centre de Recherche Paul-Pascal (CRPP, UMR 5031),  115 Avenue Schweitzer, F-33600 Pessac, France}

\author{Laura Dal Compare}
\affiliation{Dipartimento di Scienze Molecolari e Nanosistemi, 
Universit\`{a} Ca' Foscari di Venezia Campus Scientifico, Edificio Alfa, via Torino 155,30170 Venezia Mestre, Italy}

\author{Fr\'ed\'eric Nallet}
\affiliation{Univ. Bordeaux, CNRS, Centre de Recherche Paul-Pascal (CRPP, UMR 5031),  115 Avenue Schweitzer, F-33600 Pessac, France}

\author{Achille Giacometti}
\affiliation{Dipartimento di Scienze Molecolari e Nanosistemi, 
Universit\`{a} Ca' Foscari di Venezia
Campus Scientifico, Edificio Alfa,
via Torino 155,30170 Venezia Mestre, Italy}
\affiliation{European Centre for Living Technology (ECLT)
Ca' Bottacin, 3911 Dorsoduro Calle Crosera, 
30123 Venice, Italy}

\author{Eric Grelet} 
\email[]{Corresponding author: eric.grelet@crpp.cnrs.fr}
\affiliation{Univ. Bordeaux, CNRS, Centre de Recherche Paul-Pascal (CRPP, UMR 5031),  115 Avenue Schweitzer, F-33600 Pessac, France}

\date{\today}

\begin{abstract}
Charged colloids can form ordered structures like Wigner crystals or glasses at very low concentrations due to long-range electrostatic repulsions. Here, we combine small-angle x-ray scattering (SAXS) and optical experiments 
with simulations to investigate the phase behavior of charged rod-like colloids across a wide range of salt concentrations and packing fractions. At ultra low ionic strength and packing fractions, we reveal both experimentally and numerically a direct transition from a nematic to a crystalline smectic-B phase,  previously identified as a glass state. This transition, bypassing the smectic-A intermediate phase, results from minimizing Coulomb repulsion and maximizing entropic gains due to fluctuations in the crystalline structure. This demonstrates how \textit{long-range} electrostatic repulsion significantly alters 
the phase behavior of rod-shaped particles and highlights its key-role in driving the self-organization of anisotropic particles. 
\end{abstract}

\maketitle

In 1934 Eugene Wigner suggested the possibility of observing a crystalline state for 
electrons in a uniform background of positive charges at very low densities and sufficiently low temperatures \cite{Wigner1934,Wigner1938}. This 
occurs because at low densities the potential energy becomes larger than the kinetic one, thus resulting in the possibility of spatial ordering in a crystalline state. While a rigorous proof of Wigner's conjecture has been a quest for many years \cite{Giuliani2008,Zhou2021,Smolenski2021}, this concept has been extended analogously in soft condensed matter \cite{Weitz2013,Weeks2017}. Here, Wigner crystals (glasses) are assigned to any systems able to crystallize (freeze in a kinetically arrested state) at \textit{low packing fraction} and stabilized by \textit{strong long-range electrostatic repulsion} 
\cite{Lindsay1982,Leunissen2007,Russell2015,Everts2016,Angelini2014,Weeks2017,Sciortino2005}.
While the formation of ordered states by mutual repulsion between like-charged
particles has already been demonstrated \cite{Sirota1989,Monovoukas1989,Hynninen2003,Yethiraj2007,Herlach2010}, colloidal Wigner crystals and glasses additionally require an ultra
low ionic strength, so that the Debye screening length $\kappa^{-1}$, which sets the range of electrostatic repulsions, is many
particle diameters, $D$, and therefore the  
interaction between particles is nearly 
Coulombic 
($\kappa D \ll 1$)
\cite{Leunissen2007,Zaccarelli2008,Klix2010,Russell2015,Everts2016}. 

For rod-like charged particles the counterpart of a Wigner crystal would be 
 the crystalline smectic-B phase, combining lamellar organization normal to the layers and 2D long-range hexagonal order within the layers. The smectic-B phase differs from the liquid-crystalline smectic-A phase which has liquid-like order within the layers. 
Because of their shape anisotropy, 
rod-like particles display more complex phase behavior compared to spherical ones, with the formation of different liquid crystalline phases \cite{Le(ker)3&Tuinier}, from orientationally ordered nematic (N) to 
smectic-A, (SmA),  
smectic-B (SmB), and columnar (Col) phases as shown experimentally \cite{Maeda2003,Tang1995,Purdy2004,Purdy2007,Grelet2008,Kuijk2012,Grelet2014}, theoretically \cite{Onsager1949,Bohle1996,Wensink2007} and numerically \cite{Stroobants1986,Bolhuis1997,Lopes2021,Dussi2018}. 
Such a phase behavior is fundamentally 
driven by excluded volume, 
that has been extended to charged rods by introducing a charge- and ionic-strength-dependent effective diameter 
accounting for the electrical double layer thickness in case of \textit{screened} electrostatic interaction \cite{Le(ker)3&Tuinier}. 
In the opposite limit where long-range Coulomb interaction prevails 
($\kappa D \ll 1$), much less is known especially considering highly anisotropic particles 
\cite{Wierenga1998,DeMichele2007,Liu2014,Chu2020}. The complexity increases as two glass transitions associated with the translational and rotational orderings have been observed for colloidal ellipsoids both in 2D \cite{Zheng2011} and 3D \cite{Roller2021}. 
For rod-like objects, a promising system for observing a Wigner state 
is highly charged monodisperse filamentous \textit{fd} viruses, which are widely used as 
paradigm of slender particles 
in soft matter physics. 
At low densities and low ionic strength, Kang \& Dhont found experimental evidence 
\cite{Kang2013,Kang2013structural} for a Coulomb-stabilized glass transition, suggesting the Wigner \textit{glass} \cite{Bonn1999} scenario over the Wigner crystal. 

The aim of this Letter is threefold: (1) present conclusive experimental evidence regarding the 
phase behavior of \textit{fd} virus particles in the 
regime of very low salt concentrations and low packing fraction, (2) unambiguously show that, under these conditions, the system \textit{crystallizes} from the nematic phase, contrary to previous suggestions pointing toward a (Wigner) glass, (3) support these findings with simulations of rod-like charged particles interacting \textit{via} 
 Coulomb potential with tunable screening.



\begin{figure}
	\includegraphics[width=0.75\columnwidth]{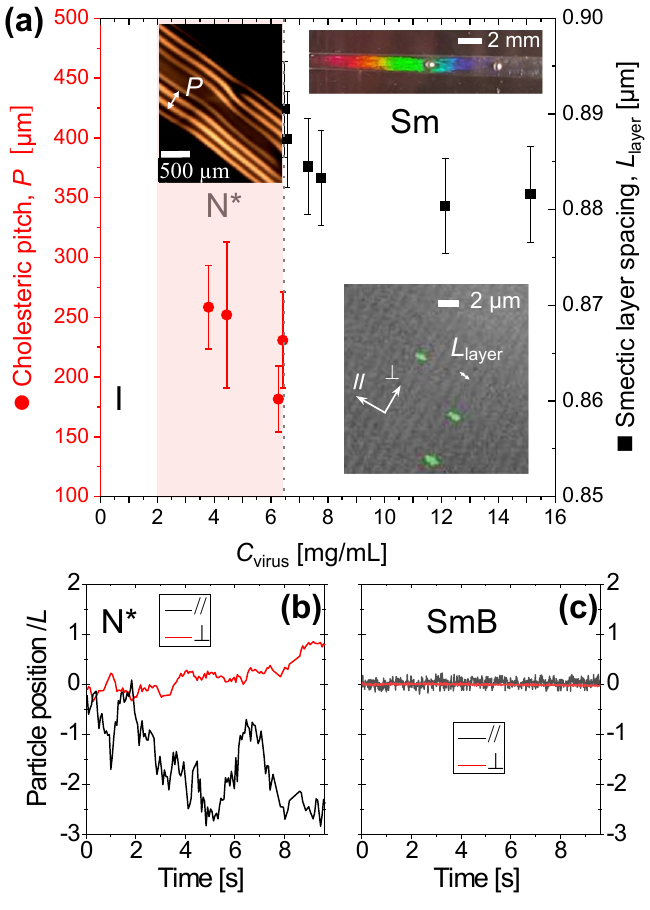}
	\caption{(a) Optical microscopy experiments revealing the phase behavior of \textit{fd} virus suspensions at low ionic strength ($I_s=0.16~$mM). Following the isotropic (I) liquid phase, the typical fingerprint texture observed between crossed polarizers (left top inset) characteristic of chiral nematic (N*) 
    phase can be identified, from which the helical cholesteric periodicity or pitch $P$ can be measured (red dots in the light red area). With increasing further virus concentration, a smectic phase forms 
    with a layer spacing $L_{\mathrm{layer}} 
    \simeq 1~\mu$m measured by differential interference contrast microscopy (bottom right inset). The micrometer-scale periodicity (full black squares) of the lamellar ordering results in Bragg reflections in the visible range making the sample iridescent (top right inset). (b) and (c): The dynamics of the system can be probed by single-particle tracking, thanks to a few labeled viral rods 
    as illustrated in the bottom right inset in (a). 
    Anisotropic diffusion is evidenced in the N* phase in contrast to the lamellar phase where the dynamics is nearly frozen with no detectable motion 
    indicating a SmB ordering. 
}
	\label{Optics}
\end{figure}

In our experiments, we made use of filamentous bacteriophages, \textit{i.e.} \textit{fd} viruses, as 
a paradigmatic example of charged rod-like colloids.   
These chiral biological objects are monodisperse, long (contour length $L=880$~nm), thin (diameter $D=7$~nm), and rather stiff (persistence length $L_p \simeq 3 L$) 
particles. The charge of \textit{fd} virus is not fixed, but regulated as the 
proteins forming the viral capsid possess pH-sensitive amino-acids 
\cite{Passaretti2020}. Such a biological system has often been used for studying the phase behavior of suspensions of charged rods at high ionic strengths ($I_s > 1$~mM). 
At physiological pH above the isoelectric point $pI_E$~4.2, \textit{fd} viruses carry a bare negative charge of about $Z_{\mathrm{virus}} \simeq 10000$ \cite{Zimmermann86}. Neglecting   charge 
condensation 
\cite{Grelet2014}, 
this leads to counterion concentrations in the millimolar (mM) range for virus suspensions self-organized into liquid crystalline phases 
(Fig.~\ref{Optics}). Therefore, to maintain a \textit{constant} and \textit{low} ionic strength -- for which the added salt is not in excess with respect to the counterions -- while simultaneously varying the virus concentration, we worked in the semi-grand canonical ensemble: \textit{each virus sample at a given concentration} is extensively dialyzed against a CO$_2$-equilibrated, TRIS-HCl solution at pH 6.5 and ionic strength 0.16~mM 
\cite{Kang2007}. 
This process takes a few weeks 
to reach equilibrium as shown in Supplemental Material \cite{SM} and results in a thick electric double layer for virus particles with a Debye screening length of $\kappa^{-1}=24~$nm. The mutual electrostatic repulsion between \textit{fd} viruses is therefore long-range, with $\kappa D \ll 1$. When faster dialysis is performed  combined with the dilution or concentration of the sample, this results in \textit{out-of-equilibrium} samples exhibiting time evolution, close to those reported by Kang \& Dhont \cite{Kang2013,Kang2013structural}.       
 \\

\begin{figure*}
	\includegraphics[width=1.75\columnwidth]{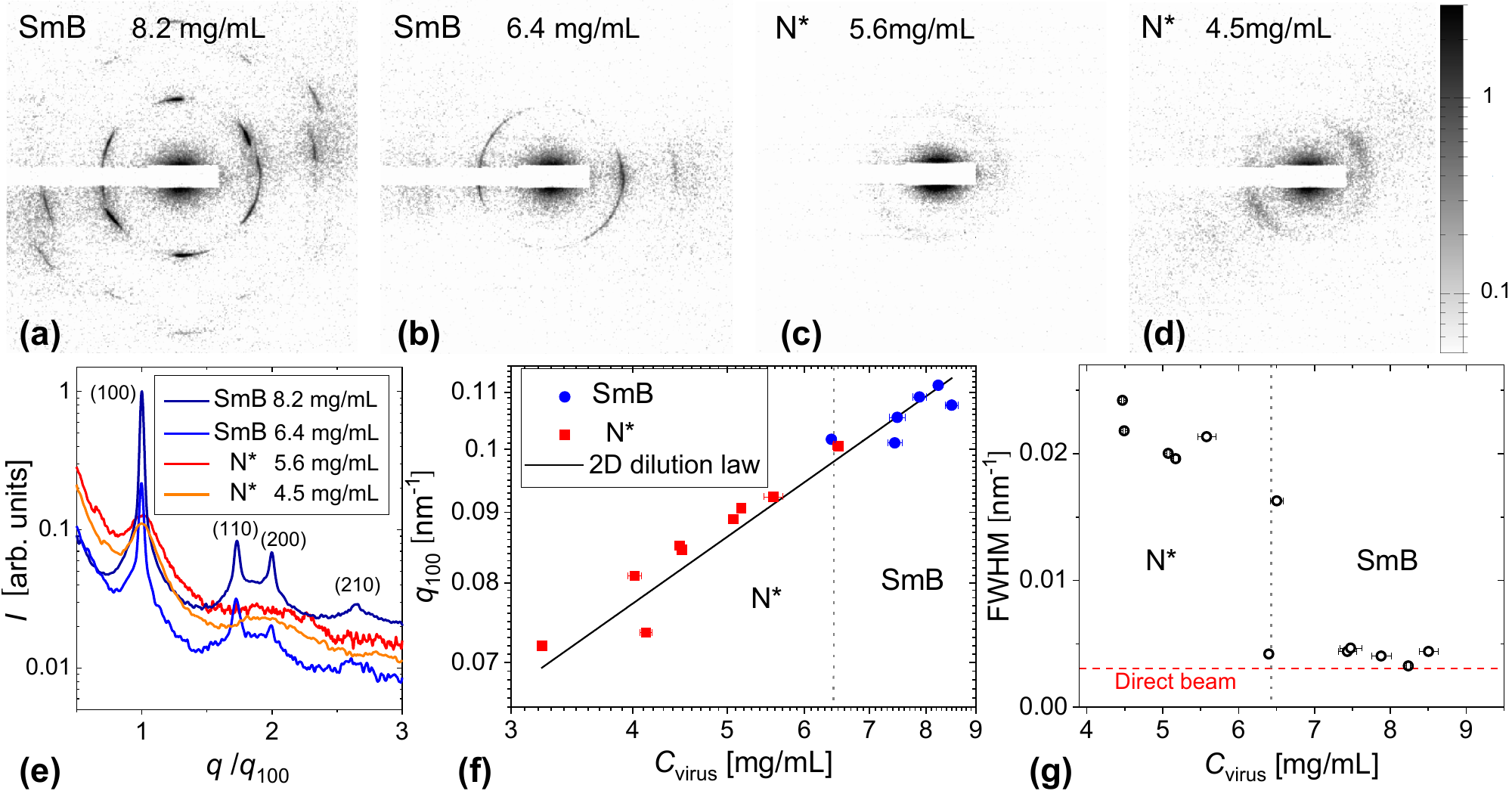}
	\caption{(a)-(d): X-ray scattering patterns probing the in-plane order normal to the long-rod axis 
    at different virus concentrations (constant ionic strength $I_s=0.16~$mM), and (e) the corresponding scattered intensities as a function of the normalized scattering wavevector $q$. The high-order Bragg reflections 
    are characteristic of long-range hexagonal order within the smectic layers, which is the signature of a smectic-B phase for the highest concentrations, \textit{i.e.} for $C_{\mathrm{virus}} > 6.4$~mg/mL as shown in (e) and (f). (f): The dilution behavior of virus suspensions follows a 2D swelling law
    ($q_{100} \propto C_{\mathrm{virus}}^{1/2}$),
    observed across all probed concentrations \cite{Grelet2016} and also in simulations (see Supplemental Materials \cite{SM}). In the (chiral) nematic phase, liquid-like ordering is evidenced by broad reflections and a high full width at half-maximum (FWHM, (g)), which is inversely proportional to the translational correlation length. This contrasts with the crystalline ordering observed in SmB which is only limited by the instrumental resolution provided by the direct x-ray beam 
    (g).   
}
	\label{SAXS}
\end{figure*}

By increasing the viral rod concentration $C_{\mathrm{virus}}$ above the isotropic liquid phase, a chiral nematic (N*) phase first appears as distinguished by its characteristic birefringent fingerprint texture observed through polarizing microscopy (Fig.~\ref{Optics}(a)). 
The helical periodicity, or cholesteric pitch $P$, can be measured and falls within the range of approximately 150 to 300~$\mathrm{\mu}$m, consistent with other studies conducted at higher ionic strengths 
\cite{Grelet2024}. Since the cholesteric periodicity far exceeds the other structural length scales in the system, our rod suspension can be approximated as a nematic phase when considering its free energy 
\cite{Purdy2004}. 
Upon a further increase in virus concentration, a change of optical texture occurs, as noticed by Kang \& Dhont 
under similar ionic conditions \cite{Kang2013,Kang2013structural}. By employing an optical objective with high numerical aperture 
(Olympus 100× PlanAPO NA=1.4), a distinct striped pattern can be discerned, characterized by a periodicity $L_{\mathrm{layer}} \simeq 1~\mu$m \cite{Grelet2014}, much smaller than the helical pitch $P$. This layered structure is iridescent and corresponds to a lamellar -- or smectic 
-- organization, where viruses with a contour length of $L \simeq L_{\mathrm{layer}}$, stand normal to the layers, as evidenced by the doping of samples with dye-labeled rods (Fig.~\ref{Optics}a). Leveraging the presence of such fluorescent viruses, the dynamics of the system has been probed \textit{via} single-particle tracking, with examples of traces provided for the two ordered phases (Figs.~\ref{Optics}(b) and \ref{Optics}(c)). 
In the N* phase, a typical Brownian diffusion consistent with nematic-like ordering is observed, characterized by enhanced motion of the rods parallel to their long axis  \cite{Lettinga2005}. Conversely, in the Sm state, no significant motion can be detected 
within the limit of spatial and time resolution of the experimental setup.  
This nearly frozen dynamics contrasts with the hopping-type diffusion observed in the SmA phase \cite{Lettinga2007}, and rather suggests crystalline SmB ordering \cite{Alvarez2017}. 
It is worth noting here that the abrupt change in particle mobility observed by single particle tracking at the N*-SmB transition is 
compatible with the claim of a 
particle arrest in the nematic phase associated with a glass state, as previously reported \cite{Kang2013,Kang2013structural}. 
Indeed, Kang \& Dhont observations can be interpreted as a slowly evolving state toward the thermodynamically stable SmB phase. 
This reinterpretation of their findings, which were based on dynamic results lacking therefore structural information, aligns with our own out-of-equilibrium experiments (see Supplemental Materials \cite{SM}), which consistently exhibit a 
transition to a crystalline smectic-B phase.



A definite evidence in favor of a crystalline ordering contiguous to the nematic phase at low added salt 
is provided by structural investigations.
Small-angle X-ray scattering (SAXS) was therefore performed using synchrotron facilities (ID02 beamline, ESRF, France). As 
$C_{\mathrm{virus}}\simeq10$mg/mL 
corresponds to very dilute suspensions with volume fractions of about 1\% and thus to large inter-rod distances, the use of synchrotron radiation 
with high brilliance and high resolution is essential to underpin the nature of the in-plane order exhibited by the \textit{fd} virus colloidal suspensions. Working with a 16~keV x-ray beam at a sample-to-detector (Frelon camera, ESRF) distance of 5~m, the full range of virus concentrations was studied, as shown in Fig.~\ref{SAXS}. For the lowest particle packing fractions ($C_{\mathrm{virus}} < 6.4$~mg/mL), the x-ray scattering patterns display broad peaks (Figs.~\ref{SAXS}(c) and \ref{SAXS}(d)) characteristic of liquid-like order (Figs.~\ref{SAXS}(e) and \ref{SAXS}(g)), consistent with the nematic ordering. 
In the presence of a glass transition, the SAXS patterns would not change significantly maintaining a liquid-like ordering.
However, with a further increase in particle concentration, sharp, resolution-limited reflections appear 
(Fig.~\ref{SAXS}(g)) with multiple orders (Figs.~\ref{SAXS}(a) and \ref{SAXS}(b)). The sequence in reciprocal space relative to the position of the first Bragg peak is 1:$\sqrt{3}$:$\sqrt{4}$:$\sqrt{7}$ (Fig.~\ref{SAXS}(e)), associated with (100), (110), (200) and (210) Miller indices characterizing
a long-range hexagonal positional order in the plane normal to the rod long axis, as also confirmed by the nearly sixfold symmetry displayed by the SAXS pattern of Fig.~\ref{SAXS}(a). Combined with the presence of layers as evidenced by optical microscopy (Fig.~\ref{Optics}(a)), this demonstrates the existence of the crystalline smectic-B phase, thereby ruling out the presence of a glass state in such low salt conditions. 
The \textit{direct} N*-to-SmB transition observed here is remarkable in view of the absence of SmA phase that is usually found as an intermediate phase. 
%
%
%
A 2D dilution law applies both in the N* and SmB range (Fig.~\ref{SAXS}(f)). The inter-rod distance $d_{\mathrm{inter}}= 4 \pi / (\sqrt{3} q_{100})$ can be calculated, and when scaled by the particle diameter $D$, it results in a ratio $d_{\mathrm{inter}}/D \geq 10$, confirming the low-density regime and thus justifying the term ``Wigner crystal'' in this context.

\begin{figure}
  \centering
  \includegraphics[width = 0.9\columnwidth]{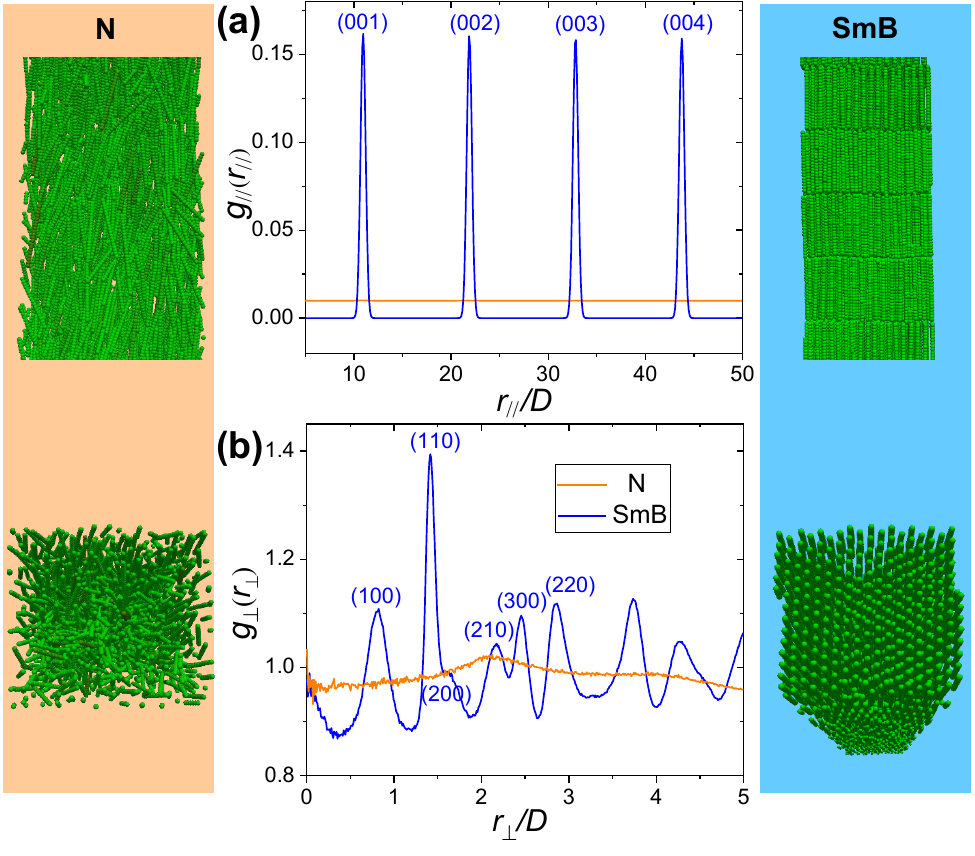}
  \caption{Parallel (a) and perpendicular (b) pair correlation functions from molecular dynamics simulations 
  at $\kappa D = 3.2$. The volume fraction has been set to $\eta=0.15$ (orange line) and $\eta=0.40$ (blue line) corresponding to the nematic 
  and smectic-B 
  phase, respectively. 
  }
  \label{fig:gr}
\end{figure}

Our experimental findings point toward a scenario where strong Coulomb repulsion, high aspect ratio, and excluded volume combine to promote the formation of a SmB phase at low volume fraction similar in spirit to the Wigner crystal predicted for the electron gas \cite{Giuliani2008}. In the colloidal realm, this phase can be considered equivalent to the Wigner body-centered-cubic (BCC) crystals observed for charged spherical colloids in the same regime, while hard spheres crystallize into a face-centered-cubic (FCC) structure \cite{Russell2015}. We now show how this scenario 
can be rationalized through molecular dynamics simulations, extending analogous numerical studies carried out for charged spheres to charged rod-like colloids \cite{Hynninen2003}.

Canonical $(NVT)$ simulations were conducted using the LAMMPS suite \cite{Plimpton1995} complemented with in-house codes for the pre- and post-processing of data. The system consists of $N=4608$ rods, with the volume $V$ adjusted to attain a desired volume fraction $\eta$. 
Simulations are initialized from a 
AAA stacking ordered configuration. 
Each rod of aspect ratio $L/D=10$ comprises 20 overlapping beads with inter-rod interactions modeled through Weeks-Chandler-Andersen (WCA) potential for the excluded volume \cite{Earl2001,DeBraaf2017,Cinacchi2017}, and Yukawa potential for the electrostatic interactions \cite{Hansen2013,Hynninen2003}. 
While matching both the range of the Debye screening lengths $\kappa ^{-1}$ and the rod surface charge density to experimental values, we tuned the volume fraction from $\eta=0.01$ to 0.65. 
Distinct phases were identified using appropriate order parameters and correlation functions \cite{Lopes2021} (see Supplemental Materials \cite{SM}). Particularly instrumental in discerning the onset of a crystalline SmB phase are the parallel $g_{\parallel}(r_{\parallel})$ and perpendicular $g_{\perp}(r_{\perp})$ correlation functions.
In the nematic phase, $g_{\parallel}(r_{\parallel})$ exhibits a flat structureless behavior, transitioning to the distinctive peaks of smectic ordering with increasing $\eta$ (Fig.~\ref{fig:gr}(a)). Likewise,  
$g_{\perp}(r_{\perp})$ displays numerous peaks (Fig.~\ref{fig:gr}(b)) associated with Miller indices akin to those observed in the SAXS experiments (Fig.~\ref{SAXS}(e)) for the SmB phase. 

\begin{figure}
  \centering
  \includegraphics[width = 0.6\columnwidth]{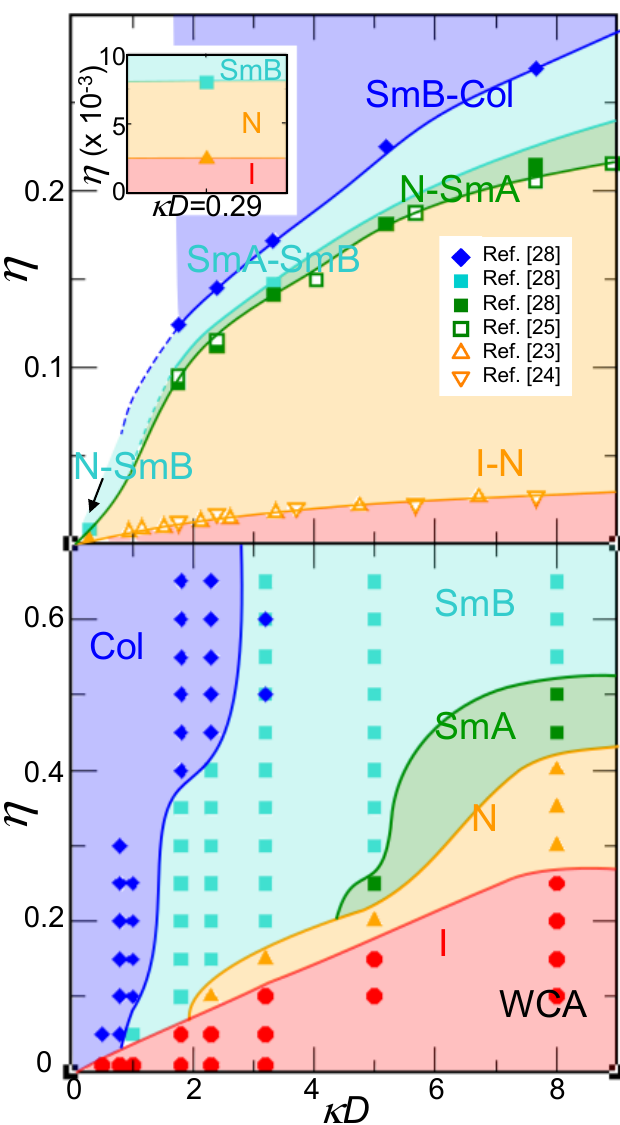} 
  \caption{\textit{Top}: Experimental phase diagram of the viral rod packing fraction $\eta$ as a function of 
  $\kappa D$ quantifying the electrostatic screening. The symbols represent the transitions between the different phases, as indicated in the graph. 
  The inset is a magnification of the transitions (I-N and N-SmB) at low ionic strength ($I_s=0.16~$mM $\iff \kappa D=0.29$). \textit{Bottom}: Calculated phase diagram 
  for charged rods exhibiting an aspect ratio of $L/D = 10$. 
  Different colors are associated with the observed thermodynamic phases: isotropic (\textrm{I}, red
  ), nematic (\textrm{N}, orange
  ), smectic-A (\textrm{SmA}, green
  ), smectic-B (\textrm{SmB}, turquoise
  ) and columnar (\textrm{Col}, blue
  ). 
  The WCA column refers to results from hard rods.
  } 
  \label{fig:experimental_phase_diagram}
\end{figure}
The numerical phase diagram 
in the $\eta$--$\kappa D$ plane 
can now be compared with the corresponding experimental one, as shown in Fig.~\ref{fig:experimental_phase_diagram}. 
In the latter, results from different experiments at higher ionic strengths ($\kappa D > 1$) are also reported in addition to those stemming from the present study, with the inset highlighting the low screening, 
low density 
behavior. While under strong screening conditions ($\kappa D \gg 1$) the behavior is primarily driven by entropy, \textit{i.e.} excluded volume, with the usual phase sequence, I-N-SmA-SmB/Col phases, the behavior at low screening conditions ($\kappa D \approx 1$) appears to be significantly different, with progressive shrinking or even disappearance of the isotropic, nematic and smectic-A phases. This leads, in simulations, to solely a crystal-like ordering (\textit{i.e.} either SmB or Col),  when $\kappa D $ is reduced below $\approx 1$. 
Remarkably, 
our simplified model of charged rods qualitatively captures 
the key features of the experimental phase diagram. Quantitative discrepancies arise primarily  
from the significant disparity in rod aspect ratio: $L/D = 10$ in simulations (chosen for computational efficiency, see Supplemental Materials \cite{SM}) vs. $L/D \approx 100$ for \textit{fd} virus. This leads to a shit in the location of phase transitions, as predicted by Onsager theory ($\eta_{I-N} \propto D/L$ for slender rods \cite{Onsager1949}). Incorporating rod flexibility could further improve quantitative agreement \cite{Chen1993,DeBraaf2017}. Crucially, simulations successfully reproduce the experimental observation of \textit{direct} crystallization into the SmB phase 
from the nematic state at sufficiently long-range Coulomb repulsion ($\kappa D \approx 2$ to 4), demonstrating the robustness and universality of this 
phase behavior, independent of specific rod details. 



Our results are 
reminiscent of calculations performed for hard-core, repulsive Yukawa
spherical particles, for which the introduction of
soft 
electrostatic interactions promote the
formation of a FCC crystal at lower volume fractions as 
$\kappa ^{-1}$ increases, as well as the onset of a new 
BCC crystalline phase that is not present in the
purely hard-sphere counterpart \cite{Hynninen2003}.  Indeed, as shown
  experimentally~\cite{Russell2015}, the soft interactions
in a colloidal Wigner crystal allow for substantial fluctuations of
particles within the crystal lattice. Albeit with a shorter
lattice spacing than the FCC lattice for the same particle volume
fraction $\eta$, BCC with only 8 nearest-neighbors becomes
entropically-favored compared to the 12 nearest-neighbors FCC for
$\eta$ low enough. Similarly, while in the smectic-B phase,
the rod fluctuations may occur differently in the longitudinal and
transverse directions relative to the layer orientation, both induce
an entropic gain compared to the alternate N or SmA 
structures. When combined with the minimization of Coulomb
repulsion due to the low-density conditions, this promotes the SmB stabilization. Additional support to this
scenario stems from the average inter-particle distance
$d_{\mathrm{inter}}$ normal to the long-rod axis, which satisfies the
condition $d_{\mathrm{inter}} > \kappa^{-1}$ in the
Wigner crystal regime, both experimentally and
numerically (see Supplemental Materials \cite{SM}) thus allowing
significant fluctuations in the 2D hexagonal lattice. 


In summary, we have investigated the experimental phase behavior of
highly charged slender rods -- filamentous \textit{fd} viruses --
under the challenging conditions of very low ionic strengths ($\kappa D
\ll 1$) and packing fractions. 
We have conclusively demonstrated a direct structural phase transition from a nematic to a crystalline smectic-B phase, irrespective of sample preparation and evolution, and proposed 
a reinterpretation of the former experiments suggesting a Wigner
glass transition
\cite{Kang2013,Kang2013structural}. Our results provide the first conclusive experimental evidence that a system of charged hard rods can form a stable crystal 
at low concentrations through strong electrostatic repulsions. 
Strikingly, our findings are well
reproduced by extensive numerical calculations of charged rods
interacting \textit{via} excluded volume and screened Coulomb
interactions. The onset of a Wigner crystalline smectic-B structure is
promoted by the entropic gain due to the large fluctuations around the
equilibrium crystal lattice position and represents the rod-like
analog of BCC Wigner crystal
obtained for charged spherical colloids \cite{Russell2015}.
Our findings significantly advance our understanding of colloidal liquid crystals in largely unexplored regions of the phase diagram, where the interplay between long-range electrostatic interactions, packing, and entropy 
leads to intriguing and unexpected phase transitions. 
This work opens pathways for exploring similar phenomena in other anisotropic systems for which the significantly modified phase behavior may be used for the rational design of novel soft self-assembled materials.

\medbreak
\begin{acknowledgments}
We thank ID02 staff and the ESRF (Grenoble, France) for the allocated synchrotron beamtime (SC-4648). A. Bentaleb is warmly thanked for his help during SAXS experiments. This project has received funding from the European Union Horizon 2020 research and innovation program under the Marie Sk\l{}odowska-Curie Grant Agreement No.~641839. A.G. acknowledges financial support by MIUR PRIN-COFIN2022 grant 2022JWAF7Y.
\end{acknowledgments}
\smallbreak
Author contributions: E.G. conceived the study; H.A., E.G., and F.N. conducted the experiments; L.D.C. and A.G. performed the simulations; E.G. wrote the manuscript with contributions from A.G.; all authors edited the paper.

\bibliography{PRL_Eric,Rods,Laura}

\end{document}